\documentclass[aps,prl,twocolumn,a4paper,10pt,notitlepage,footinbib,superscriptaddress,longbibliography]{revtex4-1}
\usepackage[english]{babel}
\usepackage{lmodern}
\usepackage[utf8]{inputenc}
\usepackage{endnotes}
\usepackage{amssymb,amsmath,amsfonts}
\usepackage{textcomp}
\usepackage{braket}
\usepackage{ulem}
\usepackage{soul}
\usepackage{graphicx}
\usepackage{dcolumn}
\usepackage{bm}
\usepackage{xcolor}
\usepackage{soul}
\usepackage{xr}

\begin{document}






\title{\textcolor{black}{Cluster size determines internal structure  
of transcription factories in human cells}}



\date{\today}

\author{Massimiliano Semeraro$^*$}
\affiliation{Dipartimento  Interateneo di  Fisica,  Universit\`a  degli  Studi  di  Bari  and  INFN,
Sezione  di  Bari,  via  Amendola  173,  Bari,  I-70126,  Italy}

\author{Giuseppe Negro$^*$}
\affiliation{SUPA School of Physics and Astronomy, University of Edinburgh, Peter Guthrie Tait Road, Edinburgh EH9 3FD, UK}
\affiliation{Dipartimento  Interateneo di  Fisica,  Universit\`a  degli  Studi  di  Bari  and  INFN,
Sezione  di  Bari,  via  Amendola  173,  Bari,  I-70126,  Italy}

\author{Giada Forte$^*$}
\affiliation{SUPA School of Physics and Astronomy, University of Edinburgh, Peter Guthrie Tait Road, Edinburgh EH9 3FD, UK}

\author{Antonio Suma}
\affiliation{Dipartimento  Interateneo di  Fisica,  Universit\`a  degli  Studi  di  Bari  and  INFN,
Sezione  di  Bari,  via  Amendola  173,  Bari,  I-70126,  Italy}

\author{Giuseppe Gonnella}
\affiliation{Dipartimento Interateneo  di  Fisica,  Universit\`a  degli  Studi  di  Bari  and  INFN,
Sezione  di  Bari,  via  Amendola  173,  Bari,  I-70126,  Italy}

\author{Peter R. Cook}
\affiliation{Sir William Dunn School of Pathology, University of Oxford, OX1 3RE, UK}

\author{Davide Marenduzzo}
\affiliation{SUPA School of Physics and Astronomy, University of Edinburgh, Peter Guthrie Tait Road, Edinburgh EH9 3FD, UK}

\email[email: ]{name@}

\begin{abstract}
\textcolor{black}{
Transcription is a fundamental cellular process, and the first step of gene expression. In human cells, it 
depends on the binding to chromatin of various proteins, including RNA polymerases and numerous transcription factors (TFs). Observations indicate that these proteins tend to form macromolecular clusters, known as {\it transcription factories}, whose morphology and composition is still debated. While some microscopy experiments have revealed the presence of {\it specialised factories}, composed of similar TFs transcribing families of related genes, sequencing experiments suggest instead that mixed clusters may be prevalent, as a panoply of different TFs binds promiscuously the same chromatin region. 
The mechanisms underlying the formation of specialised or mixed factories remain elusive. With the aim of finding such mechanisms, 
here we develop a chromatin polymer model mimicking the chromatin binding-unbinding dynamics of different types of complexes of TFs. Surprisingly, both specialised (i.e., demixed) and mixed clusters spontaneously emerge, and which of the two types forms depends mainly on cluster size. 
The mechanism promoting mixing is the presence of non-specific interactions between chromatin and proteins, which become increasingly important as clusters become larger.
This result, that we  observe both in simple polymer models and more realistic ones for human chromosomes, reconciles the apparently contrasting experimental results obtained. Additionally, we show how the introduction of different types of TFs strongly affects the emergence of transcriptional networks, 
providing a pathway to investigate transcriptional changes following gene editing or naturally occurring mutations. 
}
\end{abstract}


\maketitle

\section{Introduction}


\textcolor{black}{The 3D organization of chromatin, the filament composed of DNA wrapped around histone proteins which constitutes the building block of mammalian chromosomes, is a dynamic and intricate blueprint that is thought to be important for cellular function and gene expression~\cite{chiang2022review1}. Recent advances in microscopy and high-throughput sequencing~\cite{Kempfer2020} have revealed a rich hierarchy of 3D chromatin structures within the cell nucleus. These range from relatively small DNA loops of tens to hundreds of base pairs (bps), to large organised domains spanning over hundreds of thousands of base pairs (or kilo-base pairs, kbp), which are referred to as 
topologically-associating domains (TADs), whose segments interact more frequently among each other than with other parts of the genome.  At even larger scales, the genomic material divides into  A (active) and B (inactive) compartments, which have different gene activity and 3D compaction, whereas different chromosomes occupy distinct territories inside the nucleus 
~\cite{Pombo2015,LiebermanAiden2009,Dixon2012}.} 

\textcolor{black}{A central question in cellular biology is the extent to which this rich and multi-scale organization is influenced, or even driven, by transcription~\cite{Cook2018}, the fundamental biological process during which the information encoded in a segment of DNA is converted into RNA, to be then translated into proteins. 
On the one hand, it is widely believed that TADs remain largely invariant in cells with very different transcriptional programs (e.g., belonging to different organs) -- which points to little role for transcription in determining structure~\cite{Dixon2016} (for an opposing view, see~\cite{Rowley2017}). 
On the other hand, enzymes engaged in the process of transcription, known as RNA polymerases, tend to form aggregates inside the nucleus, often referred to as phase-separated condensates, hubs, or transcription factories~\cite{Papantonis2013,Cook2018,Cramer2019,Brackley2021,Chiang2022}. Being attached to a factory strongly enhances the transcriptional activity of a gene~\cite{Papantonis2013,Cook2018}, therefore factories are 
a primary example of a structural unit with a clear transcriptional role.}


\textcolor{black}{
A recent effective way to investigate this intricate connection between transcription and 3D chromatin structure has been provided by polymer models together with Brownian dynamics simulations~\cite{Bianco2017,dipierro2022,dipierro2018,barbi2012,bukle2018,Jost2018,jost2020,zhang2021,Natesan2021,Bianco2017,Chiariello2016,Giorgetti2014,Michieletto2016,dipierro2017}. This \textit{in silico} approach has pointed to a simple and generic mechanism -- the bridging-induced attraction or bridging induced phase separation -- that spontaneously drives formation of transcription factories~\cite{Brackley2013,Brackley2016}. Such microphase separation is due to the fact that, in this type of modelling, 
TF and polymerase complexes (TF:pol) are usually depicted as multivalent elements, so that each of them can simultaneously bind to several chromatin sites: this is reasonable as a complex of proteins can easily have more than one chromatin-binding domain. Multivalent binding triggers a positive feedback:  a TF:pol binding to the chromatin filament provokes a local increase of chromatin density as it attracts several chromatin sites. The higher chromatin density, in turn, recruits further TF:pol resulting in a cluster, or transcription factory, formation. This feedback is eventually arrested by entropic costs associated with crowding and looping more and more DNA~\cite{Marenduzzo2009,Brackley2016}.}

\textcolor{black}{The existence of clusters prompts the question: does a typical cluster mainly contain just one kind of TF, or many different ones? On the one hand, microscopy experiments suggest that different factories specialize in transcribing different sets of genes, so that any one factory typically contains mainly one type of TF. For example, active forms of RNA polymerases II and III are each housed in distinct nucleoplasmic factories that make genic and snRNA transcripts respectively~\cite{Albert2015,Cook2001,Papantonis2013}. Similarly, distinct ER$\alpha$, KLF1, and NF$\kappa$B factories specialize in transcribing genes involved in the estrogen response, globin production, and inflammation~\cite{Fullwood2009,Schoenfelder2010,Papantonis2012}. One important consequence of the formation of such specialized factories is the creation of 3D networks~\cite{pancaldi}, in which genes sharing the same TFs are co-transcribed in the same clusters.}

\textcolor{black}{
On the other hand, and in contrast to the evidence for specialized factories, 
chromatin immuno-precipitation (ChIP) techniques have revealed the existence of particular chromatin fragments, called ``highly-occupied targets'' (or HOT), which are promiscuously bound by several different TFs~\cite{Moorman2006,Foley2013,Cortini2018}. 
Additionally, single-cell transcriptional profiling
points to expression levels varying continuously as cells differentiate into other cell types, which points to a complex interplay between many factors, rather than a few acting as binary switches~\cite{Ding2022,Elmentaite2022}. Interestingly, simulations of the types described above which involve $2$ different kinds of TFs, each one binding specifically to two different TU types, show the resulting clusters often contain bound TFs of just one type, rather than mixtures, although this aspect has not been investigated in depth~\cite{Brackley2013, brackley2020b,Brackley2021, Buckle2018, Brackley2016, Brackley2017b, Bianco2017, NICODEMI201490, nicodemi2022, semeraro2022,Tiana2016,Giorgetti2014,tiana2020,Negro2023,Natesan2021}.}

\textcolor{black}{
Here, we develop a polymer model with the aim of investigating the mechanisms leading to the formation of specialised or mixed factories: i.e., clusters formed  by a single type or by multiple types of TFs respectively. Within our framework, chromatin is depicted as a polymer composed of a multicolour sequence of beads, corresponding to transcription units (TUs) of different types (or colours), each one binding to the corresponding type of transcription factors, represented as additional spheres diffusing in the system. With respect to previous works on multicolour models~\cite{Brackley2016,Bianco2018,Chiariello2016,Jost2014,Falk1,Johnstone}, there are two important differences. First, here we model chromatin transcription by making the assumption that a chromatin bead is transcribed when it binds to a TF. In this way we are able to investigate the link between 3D structure of active chromatin 
and transcription (transcriptional patterns and emerging transcriptional correlation networks), rather than solely structure as done in previous models. Second, we study the morphology of the ensuing transcriptional clusters, studying their composition and the way in which it can be affected by the 1D binding landscape, and the balance between non-specific and specific chromatin-TF interactions. } 

\textcolor{black}{
Our main result is that specialised (demixed) and mixed clusters are {\it not} mutually exclusive. More specifically, we unexpectedly find a transition, or crossover, between a specialised and a mixed clusters regime, influenced  by the size of emerging clusters. Smaller clusters are typically specialised, whereas larger clusters are more likely to be formed by different TF types. This result enables us to reconcile the apparently contrasting experimental observations cited above: it is no longer surprising that specialised and mixed clusters can coexist within the same cell, as cluster size depends, for instance, on the local 1D pattern of TU binding sites along chromatin.\textcolor{black}{It is also important to specify that, even if an investigation of a transition between small and larger transcription factories have been carried out in mouse embryonic stem cells~\cite{Cho2018}, in serum-stimulated cells~\cite{Wei2020} and in zebrafish embryos~\cite{Pancholi2021}, our study aims to provide a broader mechanistic overview.} 
We further integrated our multi-color model with experimental data, specifically DNase hypersensitive site (DHS) locations, to study human chromosomes. Here, two colours are considered as the simplest extension of the previous DHS  model  with one color~\cite{Brackley2021}, by distinguishing between  active TF:pol complex that bind respectively to cell-type-invariant and cell-type-specific TUs in strings mimicking whole human chromosomes. 
Finally, the existence of specialized and mixed factories is further validated by incorporating different types of proteins into more complex and realistic chromatin models, such as the ``highly predictive heteromorphic polymer model'' (HiP-HoP model),  which accounts for loop extrusion by cohesin-like complexes, the presence of inactive or silenced chromatin, and chromatin heteromorphism~\cite{Buckle2018}.}


\section{RESULTS}

\begin{figure}[]
\includegraphics[width=0.95\columnwidth]{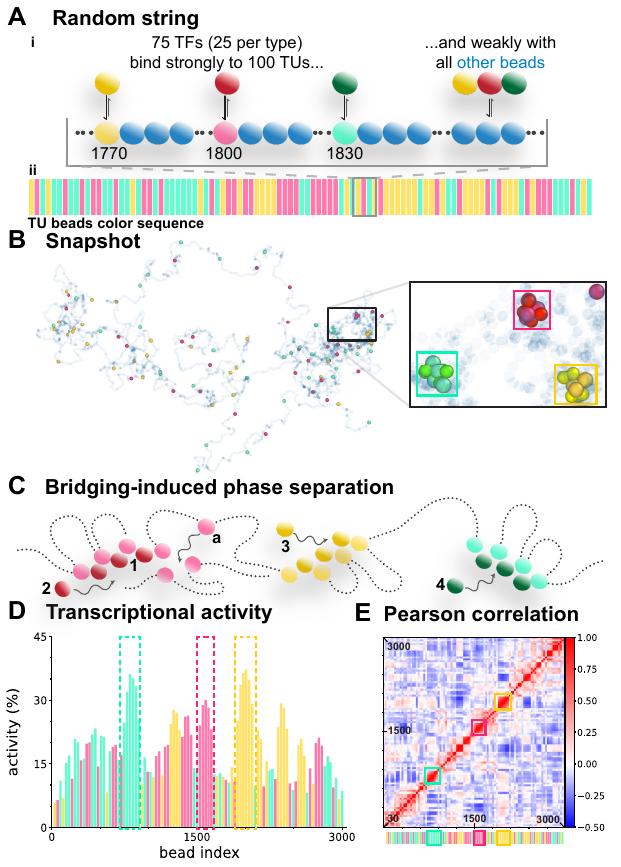}
\caption{\textbf{Toy model, with TUs coloured randomly (the random string)}. $\textbf{(A)}$ Overview. (i) Yellow, red, and green TFs ($25$ of each colour) bind strongly (when in an on state) to $100$ TUs beads of the same colour in a string of $3000$ beads (representing $3$ Mb), and weakly to blue beads. TU beads are positioned regularly and coloured randomly, as indicated in one region of the string. TFs switch between off and on states at rates $\alpha_{off}=10^{-5}~\tau_B^{-1}$ and $\alpha_{on}=\alpha_{off/4}$ ($\tau_B$ Brownian time, which one can map to $0.6-6~10^-3~s$, see SI). (ii) The sequence of bars reflects the random sequence of yellow, red, and green TUs (blue beads not shown). \textbf{(B)} Snapshot of a typical conformation obtained after a simulation (TFs not shown). Inset: enlargement of boxed area. TU beads of the same colour tend to cluster and organize blue beads into loops. $\textbf{(C)}$ Bridging-induced phase separation drives clustering and looping. Local concentrations of red, yellow, and green TUs and TFs might appear early during the simulation (blue beads not shown). Red TF $1$ -- which is multivalent -- has bound to two red TUs and so forms a molecular bridge that stabilizes a loop; when it dissociates it is likely to re-bind to one of the nearby red TUs. As red TU $2$ diffuses through the local concentration, it is also likely to be caught. Consequently, positive feedback drives growth of the red cluster (until limited by molecular crowding). Similarly, the yellow and green clusters grow as yellow TF $3$ and green TF $4$ are captured. $\textbf{(D)}$ Bar heights give transcriptional activities of each TU in the string (average of $100$ runs each lasting $8~10^{5}\tau_B$). A TU bead is considered to be active whilst within \textcolor{black}{$2.24\sigma\sim 6.7 \times 10^{-8}m$} of a TF:pol complex of similar colour. Dashed boxes: regions giving the 3 clusters in the inset in $\textbf{(B)}$. $\textbf{(E)}$ Pearson correlation matrix for the activity of all TUs in the string. TU bead number (from low to high) is reported on axes, with pixel colour giving the Pearson value for each bead pair (bar on right).  Bottom: reproduction of pattern shown in $\textbf{(A,ii)}$. Boxes: regions giving the 3 clusters in the inset in $\textbf{(B)}$.}
\label{fig1}
\end{figure}

\textbf{Toy model with different transcription factors.}

To try to solve the apparently contrasting views of segregated and mixed factories we start by  introducing a new simple polymer model, where a $3$ Mbp chromatin fragment is represented by a chain of $1000$ beads (each $30$ nm in diameter, and corresponding to $3$ kbp). Different kinds of TU beads are positioned regularly in this string, but are coloured randomly yellow, red, or green (we refer to this case as the random string). Different kinds of TFs are modelled as diffusing spheres, at first approximation of the same size of chromatin beads (see later for simulations changing the TFs size), which bind reversibly and strongly to beads of the same colour, and weakly to all others (see Fig. \ref{fig1}A, and Methods for further details). After running a Brownian-dynamics simulation, ~Fig. \ref{fig1}B shows a typical 3D conformation found in the steady state. Remarkably, clusters of TUs and TFs with distinct colours appear and disappear spontaneously. 
Such clustering is driven by the positive feedback illustrated in Fig. \ref{fig1}C; it depends critically on TFs being able to form molecular bridges that anchor loops~\cite{Brackley2013,Brackley2016, Brackley2021}.

We now assume that the spheres represent TF:pol complexes, and make the reasonable assumption that a TU bead is transcribed if it lies within  $2.25$ diameters ($2.25\sigma$) of a complex of the same colour \textcolor{black}{(small changes in this threshold do not affect qualitatively the measured activity, as shown in Figure S11)}; then, the transcriptional activity of each TU is given by the fraction of time that the TU and a TF:pol lie close together. Fig. \ref{fig1}D reports the mean activity profile down the string \textcolor{black}{obtained combining transcriptional data from $100$ independent simulation runs each lasting $8~10^5\tau_B$}; TUs with the lowest activities are flanked by differently-coloured TUs, while those with the highest activities are flanked by similarly-coloured TUs (dashed rectangles in Fig. \ref{fig1}D). As expected, a single-colour model with the same TU placement leads to a flat activity profile (Figure~S1A). Clearly, close proximity in 1D genomic space favours formation of similarly-coloured clusters.

We next examine how closely transcriptional activities of different TUs correlate~\cite{Cohen2000}; the Pearson correlation matrix for all TUs is shown in Fig. \ref{fig1}E. \textcolor{black}{This is built in the following way: for every couple $i$-$j$ of TUs, we evaluate the Pearson correlation coefficient between the transcription of $i$ and $j$, and then we colour the $ij$ (and $ji$) pixel of the matrix accordingly.} Correlations between neighbouring TUs of similar colour are often positive and strong, resulting in square red blocks along the diagonal (coloured boxes in~Fig. \ref{fig1}E highlight the $3$ clusters shown in the zoom in ~Fig. \ref{fig1}B). This effect is again due to the self-assembly of clusters containing neighbouring TUs of the same colour. In contrast, neighbours with different colours tend to compete with each other for TF:pols, and so down-regulate each other to yield smaller correlations. Correlations are more trivial in the single-color counterpart of Fig. \ref{fig1}, where the matrix yields only a positive-correlation band along the diagonal (Figure S1B). These results provide simple explanations of two mysterious effects -- the first being why adjacent TUs throughout large domains tend to be co-transcribed so frequently~\cite{Gilbert2004}. \textcolor{black}{The second concerns how expression quantitative trait loci (eQTLs)—genomic regions that are statistically associated with variation in gene expression levels—function.
While current models often attribute their effects to post-transcriptional regulation through complex mechanisms~\cite{Boyle2017,Brackley2021}, we have previously argued that any transcriptional unit (TU) can act as an eQTL by directly influencing gene expression at the transcriptional level~\cite{Brackley2021}. 
 Here, we observe individual TUs up-regulating or down-regulating the activity of others TUs -- hallmark behaviors of eQTLs that can give rise to genetic effects such as ``transgressive segregation''~\cite{Brem2005}. This phenomenon refers to cases in which alleles exhibit significantly higher or lower expression of a target gene, and can be, for instance, caused by the creation of a non-parental allele with a specific combination of QTLs with opposing effects on the target gene.  } 

\begin{figure*}[t]
\begin{center}
\includegraphics[width=1.75\columnwidth]{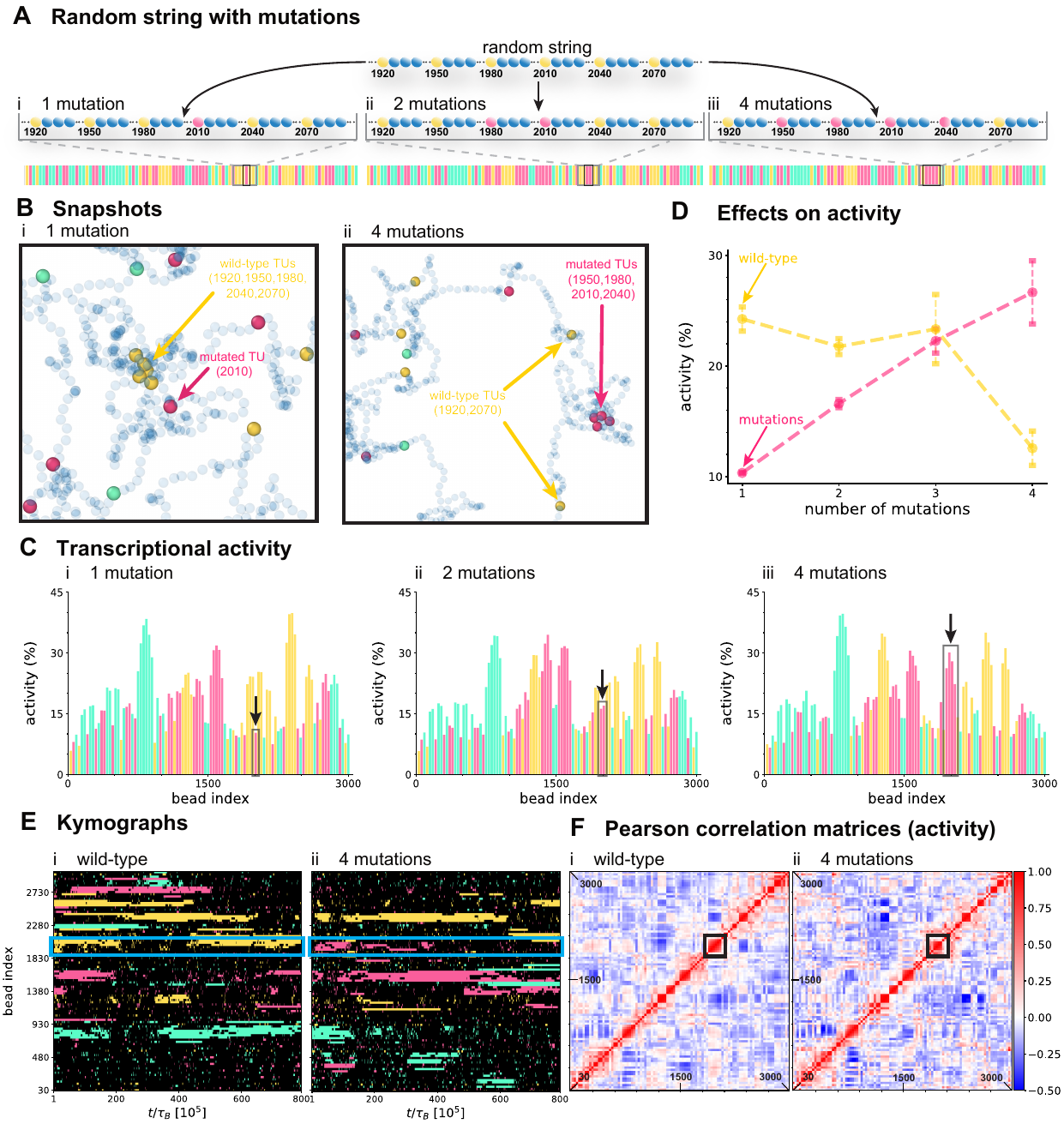}
\end{center}
\caption{\textbf{Simulating effects of mutations}. Yellow TU beads $1920$, $1950$, $1980$, $2010$, $2040$ and $2070$ in the random string have the highest transcriptional activity. $1$-$4$ of these beads are now mutated by recolouring them red. $\textbf{(A)}$ The sequence of bars reflects the sequence of yellow, red, and green TUs in random strings with $1$, $2$ and $4$ mutations (blue beads not shown). Black boxes highlight mutant locations. $\textbf{(B)}$ Typical snapshots of conformations with $\textbf{(i)}$ one, and $\textbf{(ii)}$ 4 mutations. $\textbf{(C)}$ Transcriptional-activity profiles of mutants (averages over $100$ runs, each lasting $8~10^{5}\tau_B$). Bars are coloured according to TU colour. Black boxes: activities of mutated TUs. $\textbf{(D)}$ Activities (+/- SDs) of wild-type (yellow) and different mutants. $3$ mutations: TUs $1950$, $1980$ and $2010$ mutated from yellow to red. \textcolor{black}{$\textbf{(E)}$ Typical kymographs for $\textbf{(i)}$ wild-type, corresponding to the same original string presented in Figure 1, and $\textbf{(ii)}$ $4$-mutant cases, in which 4 yellow TUs have been mutated to red.} Each row reports the transcriptional state of a TU during one simulation. Black pixels denote inactivity, and others activity; pixels colour reflects TU colour. Blue boxes: region containing mutations. $\textbf{(F)}$ Pearson correlation matrices for wild-type and $4$-mutant cases. Black boxes: regions containing mutations (mutations also change patterns far from boxes).}
\label{fig2}
\end{figure*}  

\textbf{Local mutations.} To explore the impact of introducing different colors, we characterize the effects of local mutations.  We choose the most active region in the random string -- one containing a succession of yellow TUs -- and ``mutate'' $1-4$ of these TUs by recolouring them red (Fig. \ref{fig2}A). These simulations are inspired by editing experiments performed using CRISPR/Cas9~\cite{Morgan2017}.  
Typical snapshots show red mutants are often ejected from yellow clusters (Fig. \ref{fig2}Bi), or cluster together to leave their wild-type neighbours in isolation (Fig. \ref{fig2}Bii). These changes are reflected in activity profiles (Fig. \ref{fig2}C; arrows indicate mutations). As the number of mutations in the cluster increase, activities of yellow beads in that cluster decrease (Fig. \ref{fig2}D), and new red clusters often emerge (Fig. \ref{fig2}B,ii; Fig. \ref{fig2}Ciii).


To confirm that $4$ mutations in a yellow cluster often lead to the development of a red cluster, we monitor cluster dynamics over time. \textcolor{black}{Fig. \ref{fig2}Ei illustrates a typical kymograph illustrating changes in activity of all TUs in the wild-type, corresponding to the same original string presented in Fig. 1; yellow, red, and green pixels mark activity of respective TUs, and black ones inactivity.} In this particular simulation, a yellow cluster in the region that will be mutated (marked by the blue rectangle) is present during the first quarter of the time window; it then disappears to reappear half-way through the window and then persists until the end. \textcolor{black}{From the activity profiles in Fig. \ref{fig2}C, we can observe that as the number of mutations increases, the yellow cluster is replaced by a red cluster, with the remaining yellow TUs in the region being expelled (Fig. \ref{fig2}Bii).  This behavior is reflected in the dynamics, as seen by comparing panels Ei and Eii: in the string with four mutations, transcription of the yellow TUs is inhibited in the affected region, while prominent red stripes—corresponding to active, transcribing clusters—emerge (Fig. \ref{fig2}Eii).} Pearson correlation matrices provide complementary information: the yellow cluster in the wild-type yields a solid red block indicating strong positive correlations (Fig. \ref{fig2}Fi), \textcolor{black}{while most of the pixels of this block become light-red or white} in the string with 4 mutations (Fig. \ref{fig2}Fii). 
These results confirm that local arrangements of TUs on the genetic map determine the extent to which any particular TU will cluster and so become active.


\begin{figure}[t]
\begin{center}
\includegraphics[width=1.0\columnwidth]{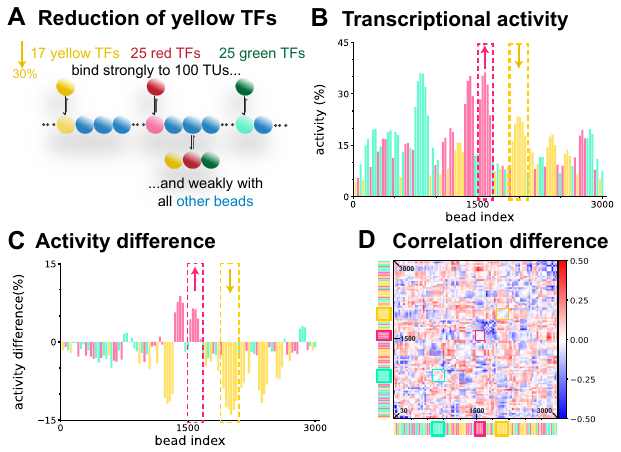}
\end{center}
\caption{\textbf{Reducing the concentration of yellow TFs reduces the transcriptional activity of most yellow TUs while enhancing the activities of some red TUs.}  \textbf{(A)} Overview. Simulations are run using the random string with the concentration of yellow TFs reduced by $30\%$, and activities determined (means from $100$ runs each lasting $8~10^{5}\tau_B$). \textbf{(B)} Activity profile. Dashed boxes: activities fall in the region containing the biggest cluster of yellow TUs seen with $100\%$ TFs, as those of an adjacent red cluster increase. \textbf{(C)}  Differences in activity induced by reducing the concentration of yellow TFs. This plot is obtained by subtracting the transcriptional activity of the wild-type, Figure~\ref{fig1}D, from that of the current system in panel B. \textbf{(D)} Pearson correlation difference matrix. This plot is obtained by subtracting the Pearson correlation matrix of the wild-type, Figure~\ref{fig1}E, from that of the current system. Boxes: regions giving the $3$ clusters from $\textbf{Figure 1B, inset}$. }
\label{fig3}
\end{figure}

\textbf{Variations in TF concentration.}
The concentration of TFs is expected to influence the global activity patterns observed and can be adjusted in our model accordingly. These simulations are motivated by experiments that reduce global TF levels using auxin-induced degrons \cite{luan2021}. Specifically, we reduce the concentration of yellow TFs binding to the random string by $30\%$ (Fig. \ref{fig3}A). As expected, transcriptional activity falls both globally and locally (see yellow dotted rectangles in Fig. \ref{fig3}B and C). Surprisingly, activity of a nearby cluster of red TUs (numbers $1080,~1110,~1170,~1200$, and $1530$ to $1650$) increases by~$50\%$ (red dotted rectangles in Fig. \ref{fig3}B and C). This effect is specific, in the sense that there is little effect on green clusters (e.g., compare Fig. \ref{fig1}D with Fig. \ref{fig3}B). We attribute this to a now-reduced steric competition for 3D space by yellow neighbours -- fewer yellow clusters are present to stunt growth of nearby red ones. 

Fig. \ref{fig3}D shows the difference in correlation between the case with reduced yellow TFs and the case displayed in Fig. \ref{fig1}E.  We can notice a change in correlation between the yellow cluster (boxed) and  its neighbour clusters, even if distant. For instance, yellow clusters are more probable to be found both turned off due to the lack of yellow TFs, and thus their activation becomes more correlated. At the same time, when yellow clusters are turned off the activation of other clusters can be affected, with a increase or decrease of correlation. 
\textcolor{black}{Overall, these results reveal numerous statistically significant correlations in gene activity both in proximal and distal regions of the genetic map. This observation aligns with the omnigenic model, which proposes that the manifestation of a genetic trait is influenced not only by the expression of core genes directly associated with the trait, but also by peripheral genes, which can exert indirect effects through gene regulatory networks. ~\cite{Boyle2017}.} 

\begin{figure}[t]
\begin{center}
\includegraphics[width=1.0\columnwidth]{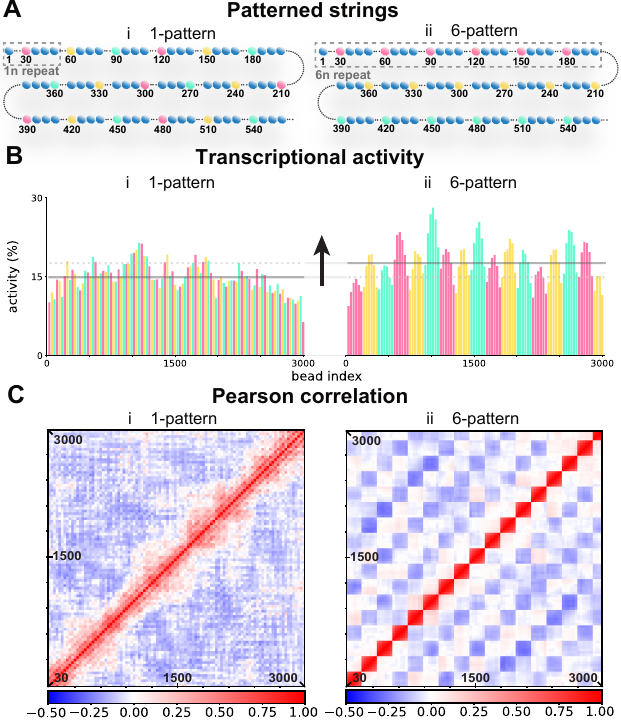}
\end{center}
\caption{\textbf{Clustering similar TUs in 1D genomic space increases transcriptional activity.} \textbf{(A)} Simulations involve toy strings with patterns (dashed boxes) repeated $1$ or $6$ times. Activity profiles plus Pearson correlation matrices are determined ($100$ runs, each lasting $8~10^{5}\tau_B$). \textbf{(B)} The $6$-pattern yields a higher mean transcriptional activity (arrow highlights difference between the two means). \textbf{(C)} The $6$-pattern yields higher positive correlations between TUs within each pattern, and higher negative correlations between each repeat.}
\label{fig4}
\end{figure}

\textbf{Effects of 1D TU patterns on transcriptional activity.}
To gain a deeper understanding of the local effects revealed by the random string, we now analyze and compare various toy strings that feature regular and repeating patterns of colored TUs (Fig. \ref{fig4} and methods for further details). Two results are apparent. First, activities (Fig. \ref{fig4}Bii) in the $6$-pattern case are higher overall (compare horizontal dotted lines), and more variable (compare activities of the two central TUs within each repeat with peripheral ones) relative to the $1$-pattern case (Fig. \ref{fig4}Bi). This is consistent with positive additive effects acting centrally within each $6$-pattern repeat, coupled to competitive negative effects of flanking and differently-coloured repeats at the edges. Second, the $6$-pattern also has a Pearson correlation matrix (Fig. \ref{fig4}Cii) that is highly-structured, with a checkerboard pattern; red blocks on the diagonal indicate high positive correlations (so the 1D $6$-pattern clearly favours 3D clustering). [Such a checkerboard pattern is not seen with a single-color model that has a correlation matrix with one red continuous diagonal when TUs are regularly spaced (Figure S1).] Additionally, blue off-diagonal blocks indicate repeating negative correlations that reflect the period of the $6$-pattern. These results show how strongly TU position in 1D genomic space affect 3D clustering and activity, and that these effects depend on inclusion of more than one colour. 

\begin{figure*}[t]
\begin{center}
\includegraphics[width=1.935\columnwidth]{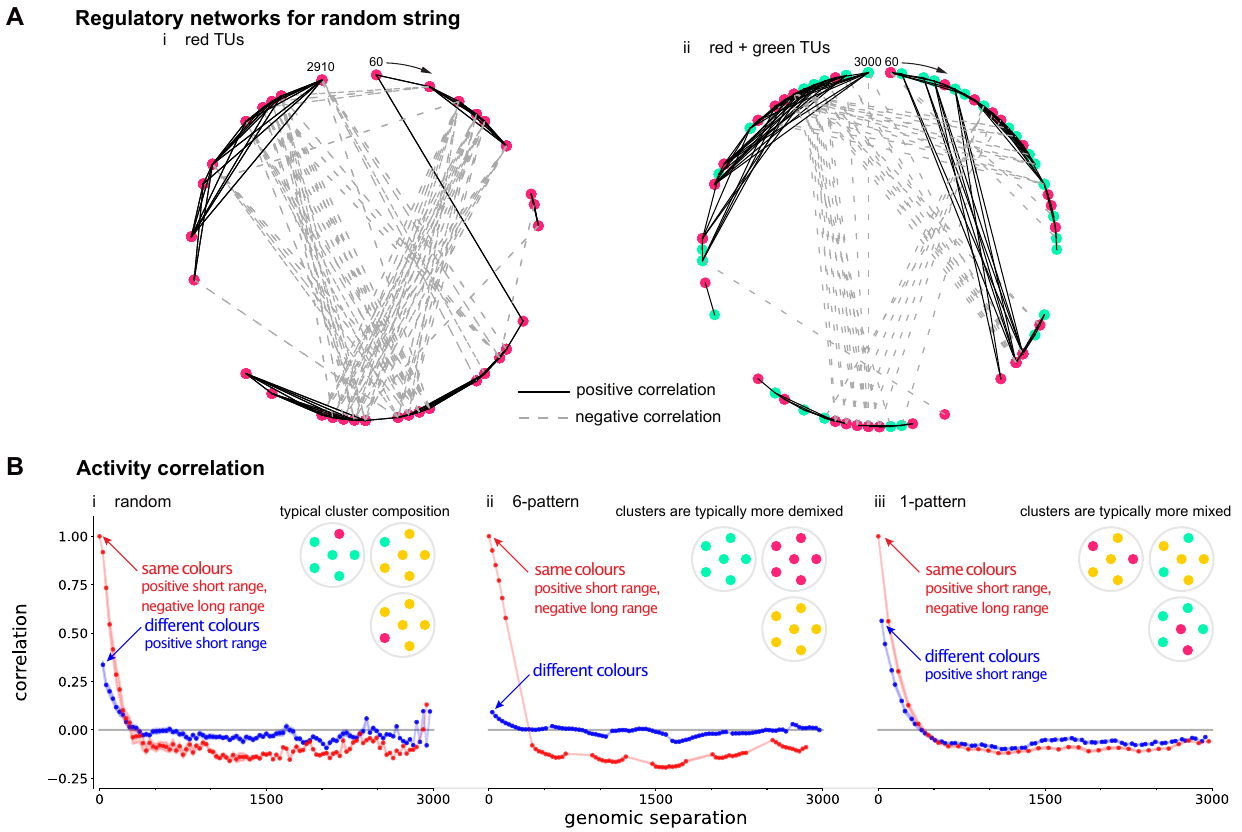}
\end{center}
\caption{\textbf{TU transcriptional networks and demixing.}  Simulations are run using the toy models indicated, and complete correlation networks {\textcolor{black}{(qualitatively reminiscent of gene regulatory networks)}} constructed from Pearson correlation matrices. \textcolor{black}{\textbf{(A)} Simplified network given by the random string. TUs from first (bead $30$) to last (bead $3000$) are shown as peripheral nodes (coloured according to TU); black and dashed grey edges denote statistically-significant positive and negative correlations, respectively (above a threshold of $0.2$, corresponding to a $p$-value $\sim5~10^{-2}$). The complete network consists of $n=100$ individual TUs, so that there are $n_c=\binom{100}{2}=4950$ pairs of TUs couples; we find $990$ black and $742$ gray edges. Since $p$-value$\cdot n_c=223$, most interactions (edges) are statistically significant. Networks shown here only correlations (i) between red TUs, and (ii) between red and green TUs.} (ii)  \textbf{(B)} Average \textcolor{black}{Pearson} correlation (shading shows +/-SD, and is usually less than line/spot thickness) as a function of genomic separation for the (i) random, (ii) $6$-, and (iii) $1$-pattern cases. Correlation values at fixed genomic distance are taken from super-/sub-diagonals of Pearson matrices. Red dots give mean correlation between TUs of the same color ($3$ possible combinations), and blue dots those between TUs of different colors ($4$ possible combinations). Cartoons depict contents of typical clusters to give a pictorial representation of mixing degree (as this determines correlation patterns); see SI for exact values of $\theta_{\rm dem}$.}
\label{fig5}
\end{figure*}

\textbf{Emergent transcriptional correlation networks.}
We have seen many positive and negative correlations between activities of TUs in the random string (Fig. \ref{fig1}). We now select significant correlations from Pearson correlation matrices (those which are $>0.2$, Fig. \ref{fig5}A) to highlight emergent interaction networks~\cite{Brackley2021}. In such networks, nodes represent each TU from first to last (other beads are not shown), and edges indicate positive (black) or negative (grey) correlations in activities of node pairs. Even for the toy random string, these networks prove to be very complex (Figure~S2A). They are also ''small-world'' (i.e., most nodes can be reached from other ones by a few steps~\cite{Watts1998,Brackley2021}). Given this complexity, we now consider simplified versions. Thus, in Fig. \ref{fig5}Ai, only interactions between red TUs are shown (the first red TU is at position $60$, the last at position $2910$, and interactions between different colours are not depicted). As expected, activities of most red TUs are positively correlated with those of nearby TUs. Conversely, negative correlations connect distant TUs, as found in the single-color model~\cite{Brackley2021}; as we have seen, binding of red TFs to any red cluster reduces the number available to bind elsewhere. 

In Fig. \ref{fig5}Aii, we consider just interactions between red TUs and green TUs. Remarkably, close-range positive correlations (black edges) are still seen between TU pairs that no longer bind TUs of the same colour. We suggest this is due to the presence of weakly-binding beads. Specifically, a red cluster organises a surrounding cloud of weakly-binding beads, and these will bind some green TFs that -- in turn -- bind green TUs. In contrast to the same-colour network in Figure~\ref{fig5}Ai, there are now more long-range positive correlations, showing that the presence of multiple colors enriches the emerging network. 

To obtain further quantitative insight into these subtle yet remarkable correlations, we compute the average of those between same- and different-colour TUs as a function of genomic separation (Fig. \ref{fig5}B). For the random string, same-colour correlations switch from clearly positive to slightly negative at about $300$ beads (Fig. \ref{fig5}Bi, red curve). Differently-coloured correlations yield a broadly-similar switch, although positive and negative values are weaker (Fig. \ref{fig5}Bi, blue curve). The $6$-pattern gives qualitatively similar trends, with the magnitude of differently-coloured correlations dampened further (Fig. \ref{fig5}Bii). In contrast, the $1$-pattern string yields largely overlapping curves (Fig. \ref{fig5}Biii). These results illustrate how the sequence of TUs on a string can strikingly affect formation of mixed clusters; they also provide an explanation of why activities of human TUs within genomic regions of hundreds of kbp are positively correlated~\cite{Hurst2004}. 

To quantify the extent to which TFs of different colours share clusters, we introduce a demixing coefficient, $\theta_{\rm dem}$ (\textcolor{black}{see Methods for definition}), which can vary between $0$ and $1$. If $\theta_{\rm dem}=1$, a cluster contains only TFs of one colour (and so is fully demixed); if $\theta_{\rm dem}=0$, it contains both red and green TFs in equal numbers (and so is fully mixed). Intuitively, one might expect $\theta_{\rm dem}$ to fall as the number of adjacent TUs of similar colour in a string fall; this is what is seen with the $6$- and $1$-patterns -- strings with the most and least numbers of adjacent TUs of similar colour, respectively (Figure~S2B; shown schematically by the cluster cartoons in Fig. \ref{fig5}B). 

Our simulations then show that in cases where same- and different-colour \textcolor{black}{Pearson} correlations \textcolor{black}{trends} overlap (as in the $1$-pattern string, \textcolor{black}{see Fig.~\ref{fig5}Biii}), clusters are more mixed (have a larger value of $\theta_{\rm dem}$). Instead, in cases where same- and different-color correlations \textcolor{black}{trends do not overlap}, or are more different (as in the $6$-pattern string, \textcolor{black}{see Fig.~\ref{fig5}Bii}), then clusters are typically unmixed, and so have a larger value of $\theta_{\rm dem}$ (Figure~S2B). \textcolor{black}{These results on activity correlation and TF cluster composition suggest that, if eQTLs act transcriptionally as expected~\cite{Brackley2021}, down-regulating eQTLs are likely to be located further from their target genes than up-regulating ones.  In addition, it is important to note that mixing is promoted by the presence of weakly binding beads; replacing these with non-interacting ones leads to increased demixing and a reduction in long-range negative correlations (Figure S3). More generally, our findings indicate that the presence of multiple TF colors offers an effective mechanism to enrich and fine-tune transcriptional regulation.}

\begin{figure*}[t]
\begin{center}
\includegraphics[width=2.0\columnwidth]{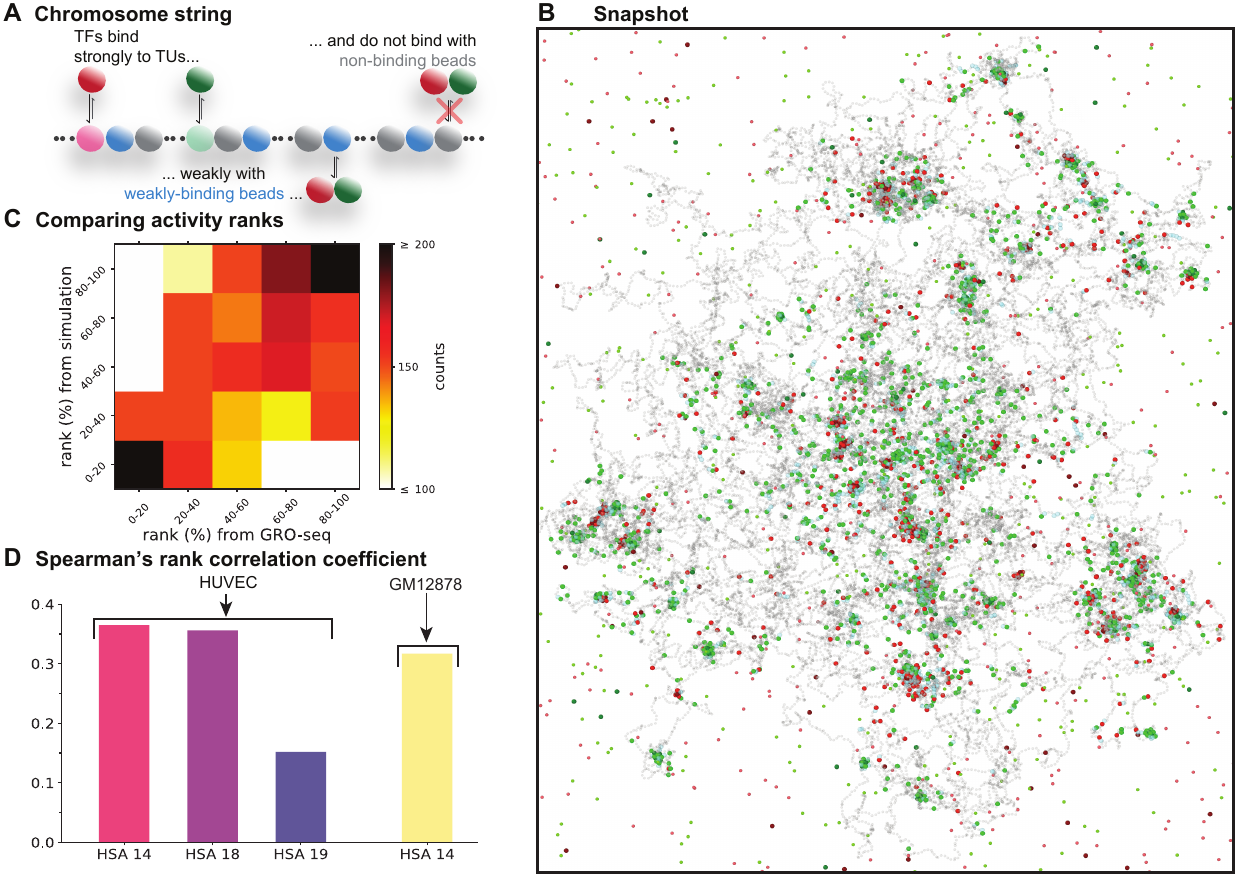}
\end{center}
\caption{\textbf{Comparison of transcriptional activities of TUs on different human chromosomes determined from simulations and GRO-seq.} \textbf{(A)} Overview of panels (A-C). The $35784$ beads on a string representing HSA14 in HUVECs are of $4$ types: TUs active only in HUVECs (red), ``house-keeping'' TUs -- ones active in both HUVECs and ESCs (green), ``euchromatic'' ones (blue), and ``heterochromatic'' ones (grey). \textcolor{black}{Red and green TFs bind to TUs of the same colour with strong (specific) interactions, while they experience a weak (non-specific) attraction to  euchromatin. Interactions between both red and green TFs and heterochromatin are purely repulsive.} \textbf{(B)} Snapshot of a typical conformation, showing both specialized and mixed clusters. \textbf{(C)} TU activities seen in simulations and GRO-seq are ranked from high to low, binned into quintiles, and activities compared.  \textbf{(D)} Spearman’s rank correlation coefficients for the comparison between activity data obtained from analogous simulations and GRO-seq for the chromosomes and cell types indicated.} 
\label{fig6}
\end{figure*}

\begin{figure*}[t]
\begin{center}
\includegraphics[width=2.0\columnwidth]{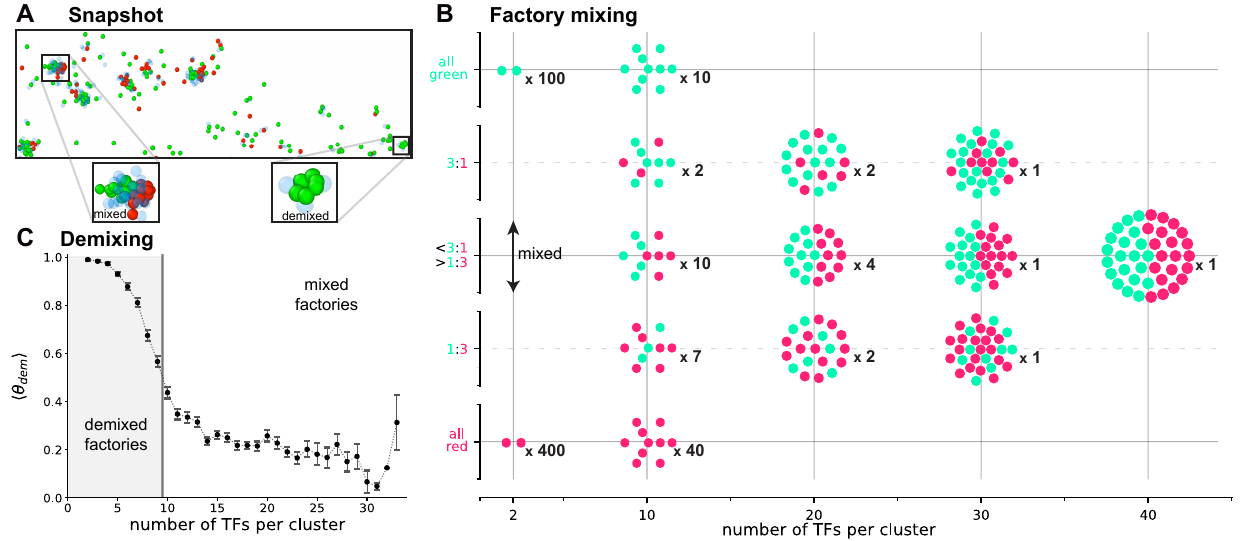}
\end{center}
\caption{\textbf{Small clusters  tend to be unmixed, large ones mixed.} After running one simulation for HSA $14$ in HUVECs, clusters are identified.
\textbf{(A)} Snapshot of a typical final conformation (TUs, non-binding beads, and TFs in off state not shown). Insets: a large mixed cluster and a small demixed one. \textbf{(B)} Example clusters with different numbers of TFs/cluster ($2$, $10$, $20$, $30$, $40$) chosen to represent the range seen from all-red to all-green (with $3$ intervening bins). Black numbers: observed number of clusters of that type seen in the simulation. \textbf{(C)} Average of the demixing coefficient $\theta_{\rm dem}$  (error bars: SD), {\textcolor{black}{ showing a crossover between demixed and mixed clusters with increasing cluster size}}. Values of $1$ and $0$ are completely demixed and completely mixed respectively. Grey area: demixed regime where $\theta_{\rm dem}$ is $>0.5$.} 
\label{fig7}
\end{figure*}

\textbf{Transcriptional activity and comparison with real human chromosomes} 
We now simulate human chromosome $14$ (HSA $14$) in HUVECs, with individual beads in the string coloured appropriately (Fig. \ref{fig6}A). Thus, TUs transcribed uniquely in HUVECs are coloured red, housekeeping TUs (i.e., ones also expressed in a stem cell, namely H1-hESCs) are green, euchromatic regions blue, and heterochromatic ones grey. Fig. \ref{fig6}B shows a typical snapshot; red and green clusters again form spontaneously. We next determine transcriptional activities, rank them in order from high to low, and compare binned rank orders with those obtained experimentally by GRO-seq (Fig. \ref{fig6}C); most counts lie along the diagonal, meaning there is a good agreement between the two data sets. More quantitatively, Spearman's rank correlation coefficient is $3.66~10^{-1}$, which compares with $3.24~10^{-1}$ obtained previously using a single-colour model ~\cite{Brackley2021}. 
In both cases the estimated uncertainty is of order $10^{-3}$ (mean and SD obtained using the bootstrap technique over $100$ trials) \textcolor{black}{and the $p$-value is $<10^{-6}$ ($2$-sided t-test), showing statistical significance.} 


Activity predictions are also improved compared to the one-colour model with HSA $18$ and HSA $19$ in HUVECs, plus HSA $14$ in GM12878 (Figure 6D and Figure S4). However, Spearman's rank coefficient for gene-poor HSA $18$ is about twice that for gene-rich HSA $19$; this may be due to additional regulatory layers in regions with high promoter density. These results show 
that our multicolour polymer model generates strings that can mimic structures and functions found in whole chromosomes.  Additionally, simulated contact maps  show a fair agreement with Hi-C data (Figure S5), with a Pearson correlation $r\sim 0.7$ ($p$-value $<10^{-6}$, 2-sided t-test). {\textcolor{black}{However, because this 2-color model does not include heterochromatin-binding proteins and cohesin-mediated active loop-extrusion, as in the Hip-Hop model later discussed}, we should not  expect a very accurate reproduction of Hi-C maps.}  
\begin{figure*}[ht]
\begin{center}
\includegraphics[width=2.0\columnwidth]{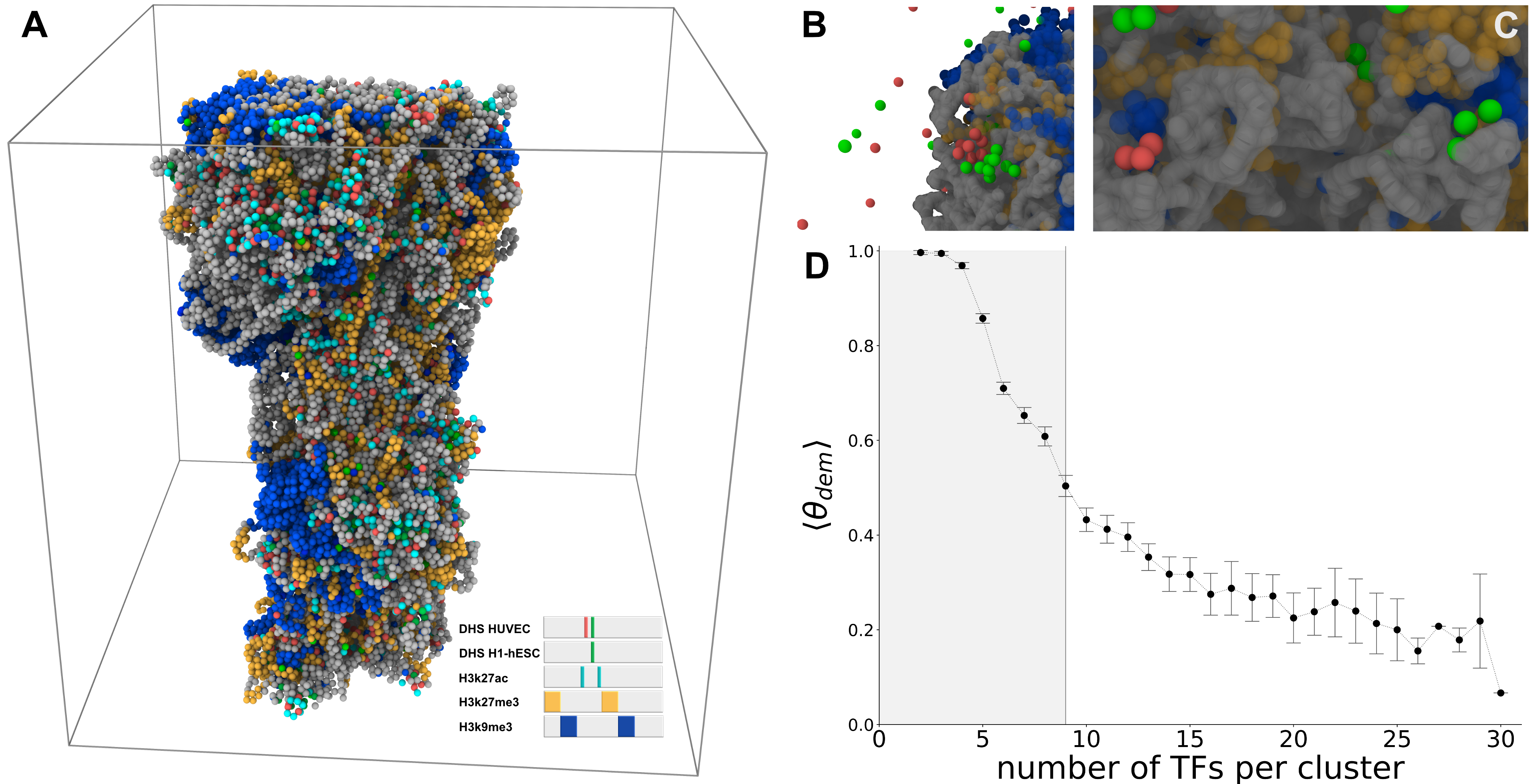}
\end{center}
\caption{\textbf{HiP-HoP model simulations: small clusters  tend to be unmixed, large ones mixed.} \textcolor{black}{(A) Snapshot of a configuration adopted  by HSA14 in HUVECs, within the HiP-HoP model. Grey regions represent  less accessible chromatin  regions poor in H3K27ac, while cyan regions represent those enriched in H3K27ac. In addition, H3K27me3 and H3K9me3 peaks determine the chromatin binding sites for polycomb-like and heterochromatin proteins, and are represented in yellow and blue respectively. As in the previous DHS multicolour model, TUs only present in HUVEC are represented in red, while the house-keeping ones in green.  (B-C) Example of clusters of proteins: large mixed cluster (B) and a small demixed one (C). (D)Average of the demixing coefficient $\theta_{\rm dem}$  (error bars: SD). Values of $1$ and $0$ correspond to completely demixed and completely mixed clusters respectively. Grey area: demixed regime where $\theta_{\rm dem}$ is $>0.5$.}} 
\label{fighip}
\end{figure*}

\textbf{Specialized and mixed clusters.} Inspection of simulation snapshots shows 1-colour clusters tend to be smaller than mixed (2-colour) ones (Fig. \ref{fig7}A). To quantify this, we count numbers and types of TFs in individual clusters (Figures~\ref{fig7}B and ~S7). Clusters with just two bound TFs never contain both colours; conversely, those with $>20$ bound TFs never contain just one colour (Fig. \ref{fig7}B). We also measure the average value of the demixing coefficient, $\theta_{\rm dem}$ (Materials and Methods). \textcolor{black}{If $\theta_{\rm dem}=1$ ($\theta_{\rm dem}=0$), this means that a cluster contains only TFs of one colour (a mixture of TFs) and so is fully demixed (maximally mixed).} {\textcolor{black}{The result is shown in Fig. \ref{fig7}C, and shows the emergence of a crossover between a demixed regime, corresponding to single-colour clusters, and a mixed regime, corresponding to multiple-colour clusters, which is triggered by an increase in cluster size. [Note that we speak of a crossover between regimes, rather than a phase transition, as clusters are finite-size and we do not consider any thermodynamic limit.]}} The cross-over point between fully mixed and demixed (where the average value of $\theta_{\rm dem}$ = 0.5) occurs when there are $\sim 10$ TFs per cluster (Fig. \ref{fig7}C): notably, this is similar to the average number of productively-transcribing pols seen experimentally in a transcription factory~\cite{Cook2018}. {\textcolor{black}{Similar results are obtained for different cell types, or chromosomes (see Figs.~S6 and S7 for the case of HSA $18$, $19$ in HUVEC, and HSA $14$ in GM12878), \textcolor{black}{and chromosomes under confinement (Fig.~S10), with realistic chromatin densities. The latter situation suggests that,  as far as the formation of transcription factories and {\textcolor{black}{the crossover between mixed and demixed clusters are concerned}}, chromatin density does not play a crucial role. Other phenomena can indeed depend on density, especially with respect to global chromatin structure (e.g., entanglements and rare long-range contacts).} \textcolor{black}{Additionally, simulations of HSA $14$ in HUVEC cells with different size of TF:pols ($0.5 \sigma$ and $0.16 \sigma$) also lead to similar results (see Fig.~S9)}. Importantly, the critical number of TFs per cluster separating demixed and mixed cluster is around $\sim 10$ in all these different cases. These results suggest that neither the sequence of TUs and its ratio to TFs (which varies among chromosomes, as for instance HSA $18$ and HSA $19$ are gene poor and gene rich respectively), nor the chromatin density affect the nature of the {\textcolor{black}{crossover}} between the regimes of demixed and mixed clusters.}} 

The {\textcolor{black}{existence of a crossover}} between specialized (demixed) and mixed factories with increasing size is therefore a generic feature of our model, and it can be explained by the following physical argument depending on non-specific binding.  Two red TFs in an unmixed cluster might stabilise $3$ loops, and so bring into close proximity only a few non-specific binding sites that could bind a green TF. In contrast, $10$ red TFs in a cluster will stabilise many loops that inevitably bring into close proximity many non-specific binding sites -- and this makes it highly likely that some green TFs will also bind nearby to create a mixed cluster. The mixing crossover provides a way to reconcile observations that some clusters are unmixed (like factories rich in polymerases II and III), and others highly mixed (like HOTs). {\textcolor{black}{This is because clusters in a single cell are generally polydisperse, or differ in size (e.g., due to the local chromatin environment, or the patterning of TUs along the genome), hence mixed and specialised factories can coexist in the same nucleus.}} Note that cluster size is a key parameter because it strongly affects the balance between non-specific and specific chromatin-protein interaction. 


Finally, as for the toy model, the balance between mixing and demixing determines correlation patterns. For example, activity patterns of same- and differently-colored TUs in the whole chromosome (Figure~S8) are much like those in the $1$-pattern model (Fig. \ref{fig5}Biii). We attribute this to $\sim 78\%$ TFs being in mixed clusters ($\theta_{\rm dem}<0.5$), and so inevitably the resulting interactions will dominate the pattern seen. 

\textcolor{black}{Our model is already inherently out of thermodynamic equilibrium, as it includes non-equilibrium switching between binding and non-binding states for chromatin-binding proteins, resembling ATP-dependent post-translational modification of such proteins\cite{Brackley2017}. There are though important principles of chromatin organisation which the presented model does not consider. First, an important remaining question is whether other active (ATP-consuming) processes naturally present in the nucleus, such as loop extrusion~\cite{Fudenberg2016}, affect the results we found. Second, following the same logic behind the multicolor polymer model presented here, it is interesting to ask whether the presence of additional types of inactive, as well as active,  TFs and chromatin beads changes the picture.} 

To answer these questions, we have turned to a more complex framework, and to the HiP-HoP model, which includes loop extrusion by cohesin-like
complexes and chromatin heteromorphism~\cite{Buckle2018,chiang2022review2}, as well as accounting for inactive, as well as active, chromatin and TFs. 
Specifically, we performed simulations for HSA14 in HUVEC  using a multicolor version of the HiP-HoP model (see SI for more details). A typical configuration is shown in Figure 8A, where grey regions represent  locally compact regions (which are poor in H3K27ac), while cyan regions represent disrupted regions (which are enriched in decompacted chromatin and in  H3K27ac). In addition,   H3K27me3 and H3K9me3 data were used to determine the chromatin binding sites for polycomb-like and heterochromatin-associated proteins (such as HP1): these are represented in yellow and blue respectively. As in the previous DHS model, TUs only present in HUVEC are represented in red, while the house-keeping ones in green.   
Inspection of simulation snapshots shows the presence of small clusters that are demixed (Fig 8C) and large cluster that are mixed (Fig. 8B). We also measure the average value of the demixing coefficient, $\theta_{\rm dem}$ (Fig. 8 D). As in the simpler DHS model, the crossover point between fully mixed and demixed (where the average value of $\theta_{\rm dem}$ = 0.5) occurs when there are $\sim 10$ TFs per cluster. These simulations further confirm the robustness and generality of our results regarding the {\textcolor{black}{mixing-demixing crossover}} between specilized and mixed transcription factories.



\section{DISCUSSION AND CONCLUSIONS}
\textcolor{black}{In summary, in this paper we have used coarse-grained simulations to study the 3D structure of human chromatin, its transcriptional dynamics, and their mutual relationship. Unlike previous works \cite{Brackley2021, Brackley2016}, here we adopt a multicolour model, viz a polymer model in which chromatin interacts with different types (colours) of complexes between polymerases and chromatin-binding transcription factors (TF:pols). This accounts for the important biological fact that most eukaryotic cells show different kinds of RNA polymerases and a variety of chromatin-binding proteins, with different biological scopes. Our model yields a number of experimentally relevant results. }

\textcolor{black}{First, we characterise the morphology of transcription factories (or clusters), arising in our model through the bridging-induced attraction~\cite{Brackley2013}. When these clusters are small, they typically contain TFs of just one colour; these are reminiscent of the specialized transcription factories found in the nucleoplasm that contain active forms of just pol II or pol III -- but not both~\cite{Xu2008a}. Instead, when factories are large, they are typically mixed (Fig. \ref{fig7}C); this provides a mechanistic basis for the formation of HOTs, where many different TFs bind promiscuously and weakly to segments of open chromatin that are often devoid of high-affinity specific binding sites~\cite{Moorman2006,Foley2013,Cortini2018}. }
The existence of a {\textcolor{black}{transition (more precisely, a crossover)}} between demixed and mixed clusters dependent on cluster size is robust to changes in TF:pol size (Fig.~S9), chromatin density (Fig.~S10) and the inclusion of active process such as loop extrusion, incorporated through the HiP-HoP model (Fig.~8). The latter simulations also show that the existence of a demixing transition (and the critical size threshold) is not affected by other structurally important ingredients in the model such as the presence of silence or inactive chromatin, and chromatin heteromorphism~\cite{Buckle2018}. \textcolor{black}{This confirms that the mechanism behind the transition is the shift in balance between non-specific and specific chromatin-protein interactions: the former becomes more dominant as cluster size increases.  Interestingly, the mechanisms that determine whether a gene belongs to a specialised or mixed factory remain unclear~\cite{Razin2011}.} \textcolor{black}{However, our results suggest that the TF cluster size, along with the 1D TU patterning along the chromatin filament, plays a crucial role, as it links  3D chromatin structure, transcription factory morphology, and gene expression. Specialised and mixed factories thus  emerge naturally from TUs arrangement, without the need for additional ingredients, such as post-transcriptional biochemical regulation. }

\textcolor{black}{
We propose that the predicted demixing–mixing crossover may be experimentally testable in the future through techniques capable of detecting multi-way chromatin interactions, such as SPRITE and GAM~\cite{Kempfer2020}. At present, however, these methods have primarily enabled the detection of three-way chromatin contacts~\cite{Quinodoz2018, Quinodoz2021, Beagrie2017, Beagrie2023}, and statistical data on higher-order interactions remain limited or unavailable~\cite{Liu2021}. Nevertheless, it is reasonable to expect that ongoing advancements in these technologies will yield increasingly comprehensive datasets, potentially allowing for direct experimental validation of our predictions.}


Second, we see remarkable positive and negative correlations in the transcriptional activities of different TUs. For example, activities of same-colour and nearby TUs tend to be strongly positively correlated, as such TUs tend to co-cluster (Fig. \ref{fig5}). \textcolor{black}{Conversely, activities of similar TUs lying far from each other on the genetic map are often weakly negatively correlated, as the formation of one cluster sequesters some TFs to reduce the number available to bind elsewhere (Figure S12). }

Taken together, these results provide simple explanations of why adjacent TUs throughout large domains tend to be co-transcribed so frequently~\cite{Hurst2004}, as they are likely to gather together in the same cluster. Results also show how one eQTL might up-regulate some TUs and down-regulate others, that can lead to genetic effects like ``transgressive segregation''~\cite{Brem2005}. 

Third, we can predict effects of local mutations and genome edits that often induce distant omnigenic effects uncovered by genome-wide association studies~\cite{Boyle2017,Brackley2021}. For example, mutations that switch a binding site of one TF to another can convert a cluster of one colour into another (Fig. \ref{fig2}). Similarly, global effects of knocking down TF levels are easily assessed (Fig. \ref{fig3}). 

Fourth, we also predict transcriptional activities of all TUs (both genic and non-genic) on whole human chromosomes by including cell-type-invariant and cell-type-specific TFs (Fig. \ref{fig6}). We find this yields a better correlation with GRO-seq experimental data than a single-colour model (where just one TF binds to all TUs similarly). This result underscores the importance of including different TFs in polymer models. 

Finally, all our results point to the importance of the 1D pattern of TUs and TF-binding sites on chromosomes in determining activity. In other words, 1D location is a key feature determining transcriptional patterns, and so cell identity. We speculate this is why relative locations of active regulatory elements are so highly conserved. For instance, despite human enhancers evolving much more rapidly than their target protein-coding genes, the synteny between the two (over distances up to $2$ Mbp) is highly conserved~\cite{berthelot2018,laverre2022}. 


In the future, \textcolor{black}{it would be interesting  to incorporate the effects of transcription elongation into our model. Although the current implementations of both the simplified toy model and the more detailed HiP-HoP model show good agreement with GRO-seq data, further investigation is needed to assess how molecular motors —such as RNA polymerases—and interactions between RNA molecules and nuclear proteins, including SAF-A~\cite{Nozawa2017,Marenda2024}, contribute to the three-dimensional organization of chromatin.}
\textcolor{black}{Additionally, it would be valuable to consider the effects of hydrodynamic interactions~\cite{Eshghi20231,Eshghi2023,PhysRevXMahajan}. In our model, as in most chromatin polymer models in the literature~\cite{chiang2022review2}, hydrodynamic interactions and the resulting spatiotemporal correlations are neglected. Whilst this choice provides significant computational advantages, it also represents a limitation. 
Although passive hydrodynamic flow associated with polymer motion is likely screened inside the nucleus, the dipolar forces exerted by molecular motors may be strong enough to induce ordering in the intranuclear polymer melt~\cite{AlexandraZidovska}.  In fact, recent experiments using Displacement Correlation Spectroscopy,  used to map chromatin movements throughout the nucleus in live cells,  revealed that chromatin exhibits rapid, uncorrelated motions at short timescales  and slower, correlated motions over $\sim \mu$m domains at longer timescales~\cite{AlexandraZidovska}. 
While the typical sizes of the emerging clusters we observe is about an order of magnitude smaller than that of these domains, 
incorporating hydrodynamics 
would help elucidate the effect of coherent chromatin dynamics on cluster formation and coarsening. }

\textcolor{black}{
In addition, it would be of interest to extend the  results presented here to incorporate many more types of TFs and TUs within the framework of  HiP-HoP ~\cite{Buckle2018}, and to include dynamic epigenetic modifications~\cite{Brackley2017a, Michieletto2016,Olarte}.
From a theoretical point of view, we hope our results will stimulate the development of theories to understand the mixing-demixing crossover more fundamentally from a polymer physics view-point, as well as more work on the interrelations between 3D structure and function in chromosomes. }

\section{METHODS}

\textcolor{black}{Chromatin fibres are modelled as bead-and-spring polymers \cite{Brackley2013, brackley2020b,Brackley2021, Buckle2018, Brackley2016, Brackley2017b, Bianco2017, NICODEMI201490, nicodemi2022, semeraro2022,Tiana2016,Giorgetti2014,tiana2020,semeraro2022,Natesan2021}, where each monomer (diameter $\sigma$) represents $3$ kbp packed into a $30$ nm sphere~\cite{Brackley2013,Brackley2021,semeraro2022}. Different TFs (or TF:pol complexes) are modelled as differently-coloured spheres (also with diameter $\sigma$) able to bind (when in an ``on'' state) to cognate sites of the same colour that are scattered along the polymer. }
{\textcolor{black}{TF:pols complexes and TUs  have  the same size, since the former 
represents both transcription factors and polymerases, which in human cells are about  5 nm and 25 nm respectively.
We also simulated the case in which TF:pol size is smaller ($0.5 \sigma$ and $0.16 \sigma$, Fig. S9) to explore the potential effect of protein size.}}

Each TF and TF:pol switches between ``off'' (non-binding) and ``on'' (binding) states to reflect the post-translational modifications that occur in many TFs. Polymer beads are either non-binding (``heterochromatic''), weakly-binding (``euchromatic''), or strongly-binding (containing cognate sites). TFs bind non-specifically to all weakly-binding beads, and strongly only to TUs of the same colour. {\textcolor{black}{TUs in our model represents regulatory elements such as promoters and enhancers, and as discussed below in practice can be identified with DNase hypersensitive regions, which are very sticky for a wide range of TFs, or active protein complexes~\cite{encode}}.}

The system evolves in a cubic simulation domain with periodic boundary conditions through constant-temperature Langevin dynamics that are integrated numerically by the LAMMPS simulation package~\cite{lammps}. 

In our model, as in most chromatin polymer models in the literature~\cite{chiang2022review1},   hydrodynamic interactions are neglected. While this choice offers significant computational advantages, it also presents a limitation. Although passive hydrodynamic flow associated with polymer motion is likely screened inside the nucleus, dipolar forces exerted by molecular motors may still be strong enough to induce ordering in the intranuclear polymer melt, as discussed in \cite{AlexandraZidovska}(see also the Discussion section for additional comments on the possible effects of hydrodynamics). Averages are evaluated over $100$ independent runs for each case. The TF volume fraction of each colour is set to $\sim 3~10^{-5}$, and the polymer volume fraction to $\sim 2~10^{-3}$. We note though that the key control parameter is the ratio between the number of TFs and that of TUs, for each colour. More information about the model can be found in the Supplemental Information (SI).


Several quantities are monitored to describe the system's behavior. Mean transcriptional activity is measured as the fraction of time that a TU is ``transcriptionally active'', \textcolor{black}{i.e. within a fixed distance of a TF,} in $100$ simulations, and so represents a population average (each simulation run may be thought of as a different cell). \textcolor{black}{In order to be quite conservative
about the evaluation of the transcriptional activity for TUs in clusters, we adopt $2.25\sigma$ as transcriptional threshold. Our point of view is in fact to
consider transcriptionally active all TUs in clusters, even when TFs are a bit further. However, we also checked that lower distance thresholds do not affect evaluation of transcriptional profiles in any significant way (see Fig.S11 in SI).} Transcriptional activity is then compared with experimental data on transcriptional activity, obtained via GRO-seq -- a method providing a genome-wide average readout of ongoing transcription of both genic and non-genic TUs in cell populations \cite{Core2014,Jordan2019}. The mean transcriptional Pearson correlation between all pairs of TUs is also evaluated, and a graphical overview of this feature is provided via the Pearson correlation matrix. We also analyse clusters/factories of bound and spatially-proximate TFs, count the number of TFs of similar colour in each cluster, and introduce a demixing coefficient
\begin{equation}
    \theta_{\rm dem}=\frac{nx_{i,max}-1}{n-1},
    \label{eq:dem_coeff}
\end{equation}
where $n$ is the number of colors, \textcolor{black}{$i$ is an index denoting each of the colors present in the model and $x_{i,max}$ the largest fraction of TFs of the same $i$-th color in a single TF cluster}. If $\theta_{\rm dem}=1$, this means that a cluster contains only TFs of one colour and so is fully demixed; if $\theta_{\rm dem}=0$, the cluster contains a mixture of TFs  of all colors in equal number, and so is maximally demixed. More details can be found in the SI. 


We consider two different types of string, one with $M=3000$ beads (or $9$ Mbp) which is referred to as a ``toy'' string, and a second representing a whole human chromosome. 
Chromosomes are initialised in both cases as random walks.
{\textcolor{black} {An alternative possibility would be to start from mitotic configuration as in~\cite{Rosa2008}, which would remove entanglement in the initial condition. Experience with similar models (e.g., see~\cite{Jost2014}) suggests that a different initial condition will be important for the very large-scale structure but not for the scale at which transcriptional clusters form, which is the one we are most interested in here.}}


\textbf{Toy model.}
The toy model is built by placing one yellow, red, or green TU every $30$ weakly-binding beads, giving a total of $100$ TUs of all types in a string of $3000$ beads ~\cite{Brackley2013}. Various different sequences of TU colour down the string are considered. In one -- the ``random'' string -- TU colours are chosen randomly (see Fig. \ref{fig1}a and SI for the specific sequence generated). In a second and third -- the ``$1$-pattern'' and ``$6$-pattern'' strings -- TU colors follow a repeating pattern (red, then yellow, then green) $1$ or $6$ times (see Fig. \ref{fig4}). 
{\textcolor{black}{We made these choices for the sequences of TUs as they are useful to show how 1D patterns affect resulting cluster morphology. In this respect, these patterns in the toy model are only representative. At the same time, the ratio between TFs and TUs are close to those used below for human chromosome simulations. }}

For the random string, we monitor how the system responds to different perturbations. Local ``mutations'' are inspired by editing experiments performed using CRISPR/Cas9~\cite{Morgan2017}. One to four mutations are mimicked by switching selected yellow beads inside a cluster of consecutive yellow TUs (between TUs $1920$ to $2070$) to red ones (Fig. \ref{fig2}). Thus, conversion of TU bead $1980$ gives a string with $1$ mutation, of $1950$ and $1980$ gives $2$ mutations, of $1950$ to $2010$ gives $3$ mutations, and $1950$ to $2040$ gives $4$ mutations. Global perturbations are inspired by experiments reducing global levels of TFs using auxin-induced degrons \cite{luan2021}. Here, we study the effects of reducing the concentration of yellow TFs by $30\%$.

\textbf{Human chromosomes.}
Our reference case for whole human chromosome simulations in the main text is the mid-sized human chromosome HSA $14$ ($107$ Mbp), coarse-grained into $M=35784$ beads. For Fig. \ref{fig6}, weakly- and strongly-binding beads are identified (using ENCODE data \cite{encode} for human umbilical vein endothelial cells, HUVECs) by the presence of H3K27ac modifications and DNase-hypersensitivity sites (DHSs) in the $3$ kbp region corresponding to that bead -- as these are good markers of open chromatin and active TUs (both genic and non-genic), respectively. For Fig. \ref{fig6}, TUs are split into ones only active in HUVECs and others (``house-keeping'' ones) that are also active in H1-hESC cells (again using DHS sites and ENCODE data). Then, if a TU appears in both HUVECs and H1-hESCs, it is marked as housekeeping and coloured red; if it appears only in HUVECs it is marked as HUVEC-specific and coloured green. This allows an intuitive and simple multicolour model of HUVECs to be constructed. All remaining beads (which are not either weakly-binding or TUs) are non-binding. This approach represents a generalisation of the DHS model described in \cite{Brackley2021}, so we call it the multicolour DHS model. {\textcolor{black}{For the simulations shown in the main text TF:pols complexes and TU size is the same ($\sigma$, corresponding to 30 nm at our resolution).
This is justified by the fact that our TF:pol represents both transcription factors and polymerases. A polymerase is about 25 nm in human cells \cite{Cook2001}, while transcription factors are typically at least $5$nm in size.  
We also considered the case in which TF:pol size is smaller ($0.5 \sigma$ and $0.16 \sigma$, Fig. S9) to explore the potential effect of protein size: as we shall see, this does not qualitatively affect our conclusions and results.}}

We also consider HSA $18$ ($80$ Mbp, $26026$ beads) and $19$ ($58$ Mbp, $19710$ beads) in HUVECs, chosen as they represent gene-poor and gene-rich chromosomes, respectively. Additionally, we consider HSA $14$ in the B-lymphocyte line GM12878 (again, colours are chosen by combining DHS data for GM12878 and H1-hESCs). H3K27ac and DHS data is again from ENCODE.

The multicolor DHS model was also applied within a more realistic chromatin framework, the ``highly predictive heteromorphic polymer model'', or HiP-HoP model \cite{bukle2018}.
This is a much more sophisticated model which takes into account: (i) loop extrusion; (ii) inactive (as well as active) chromatin folding; (iii) chromatin heteromorphicity (different local compaction of chromatin according to acetylation).  More details on the HiP-HoP model are given in the SI.

For human chromosomes, transcriptional-activity data obtained from simulations and GRO-seq are compared in two ways \cite{Brackley2021}. First, we rank activities of each TU, and build a two-dimensional histogram providing an overview of the agreement between the two sets of ranks. Second, we quantify Spearman's rank correlation coefficient between numerical and experimental data (SI for more details).

\section{ACKNOWLEDGEMENTS}
M.S.  G.N. and G.F. contributed equally to this work. The work has been performed within the HPC-EUROPA3 Project (INFRAIA-2016-1-730897), with the support of the EC Research Innovation Action under the H2020 Programme.  We acknowledge funding from MIUR Project No. PRIN 2020/PFCXPE, and from the Wellcome Trust (223097/Z/21/Z).
G.F. acknowledges support from the Leverhulme Trust (Early Career Fellowship ECF- 2024-221).

\section{Data and code Availability}
All experimental data used in this paper are available in the ENCODE database~\cite{encode}.
All custom scripts used for the simulations presented here are available in the Zenodo database \url{10.5281/zenodo.17408464}.

\subsection{Conflict of interest statement.} None declared.

\bibliography{references}

\end{document}